\documentclass[%
 reprint,
 showpacs,
 amsmath,amssymb,
 aps,
]{revtex4-1}
\usepackage{amsmath}
\usepackage{cases}
\usepackage{amssymb}
\usepackage{CJK}
\usepackage{graphicx}
\usepackage{dcolumn}
\usepackage{bm}
\usepackage{color}
\usepackage[pagewise]{lineno}

\begin{document}
\begin{CJK*}{GBK}{} 

\preprint{APS/123-QED}

\title{Isobaric yield ratio difference between the 140 $A$ MeV $^{58, 64}$Ni + $^{9}$Be reactions studied by antisymmetric molecular dynamics model}
\author{C. Y. Qiao}
\author{H. L. Wei}
\author{C. W. Ma}
\thanks{Email: machunwang@126.com}
\author{Y. L. Zhang}
\author{S. S.  Wang}

\affiliation{
Institute of Particle and Nuclear Physics, Henan Normal University, \textit{Xinxiang 453007}, China
}



\date{June 29, 2015; Received 11 June 2015; Revised 19 June 2015}
\begin{abstract}
\begin{description}
\item[Background] The isobaric yield ratio difference (IBD) method is found to be sensitive to the density difference of neutron-rich nucleus induced reaction around the Fermi energy.
\item[Purpose] An investigation is performed to study the IBD results in the transport model.
\item[Methods] The antisymmetric molecular dynamics (AMD) model plus the sequential decay model GEMINI are adopted to simulate the 140$A$ MeV $^{58, 64}$Ni + $^{9}$Be reactions. A relative small coalescence radius R$_c =$ 2.5 fm is used for the phase space at $t =$ 500 fm/c to form the hot fragment. Two limitations on the impact parameter ($b1 = 0 - 2$ fm and $b2 = 0 - 9$ fm) are used to study the effect of central collisions in IBD.
\item[Results] The isobaric yield ratios (IYRs) for the large--$A$ fragments are found to be suppressed in the symmetric reaction. The IBD results for fragments with neutron-excess $I = $ 0 and 1 are obtained. A small difference is found in the IBDs with the $b1$ and $b2$ limitations in the AMD simulated reactions. The IBD with $b1$ and $b2$ are quite similar in the AMD + GEMINI simulated reactions.
\item[Conclusions] The IBDs for the $I =$ 0 and 1 chains are mainly determined by the central collisions, which reflects the nuclear density in the core region of the reaction system. The increasing part of the IBD distribution is found due to the difference between the densities in the peripheral collisions of the reactions. The sequential decay process influences the IBD results. The AMD + GEMINI simulation can better reproduce the experimental IBDs than the AMD simulation. 
\end{description}
\end{abstract}

\maketitle
\end{CJK*}


\section{introduction}

The isobaric yield ratio difference (IBD) method, which is similar to the isoscaling method \cite{HShanPRL,MBTsPRL01iso}, has been developed to study chemical potentials of neutrons and protons \cite{IBD13PRC,IBD13JPG} or the nuclear density \cite{IBD14Ca,IBDCa48EPJA,NST2015IBD} in heavy-ion collisions. Based on the isobaric yield ratio (IYR), the IBD method provides cancellation of both the system dependence parameters \cite{Huang10,Ma11PRC06IYR,Huang10Powerlaw,Huang10NPA-Mscaling}, the free energies of isobars \cite{IBD13PRC}, and the terms contributing to the free energy of fragments \cite{Huang10,WadaNst13,MA12CPL09bsbv,MaCW12EPJA,MaCW12CPL06,MaCW13CPC} in the formula determining the cross sections of fragment. The target dependence of IBD has been studied by investigating the measured fragments in 140$A$ MeV $^{40, 48}$Ca and $^{58, 64}$Ni projectile fragmentation reactions on the $^9$Be and $^{181}$Ta targets \cite{IBD15Tgt}, which have been performed  by Mocko \textit{et al} at the National Superconducting Cyclotron Laboratory (NSCL) in Michigan State University \cite{Mocko06}. In addition, the Shannon information entropy theory is also adopted to explain the IBD method for a better understanding of it \cite{info1,info2}. At the same time, IYRs has been used to extract the temperature for fragments in heavy-ion collisions \cite{Ma12PRCT,Ma13PRCT,Ma2013NST,XQNst15}.

In a typical IBD distribution, there is a plateau part plus a changing (increasing or decreasing) part as the function of the mass numbers $A$ of fragments \cite{IBD13PRC,IBD13JPG,IBD14Ca,IBDCa48EPJA}. The plateau part is explained as denoting the region where the chemical potential  (or the density) difference of neutrons and protons changes very little. The modified statistical abrasion-ablation (SAA) model \cite{MaCW09PRC} was used to study the IBD and the nuclear density difference of neutrons and protons between the calcium-isotope-induced reactions \cite{IBD14Ca,IBDCa48EPJA}, which suggests that the IBD is sensitive to the nuclear density extracted from the prefragments, but the sensitivity is weakened in the results for the final fragments. Because the SAA calculation cannot reproduce some of the IYR distributions, the IBD results by SAA cannot well explain the measured data \cite{IBD14Ca}. The SAA model has a simple collision and deexcitation mechanism \cite{FangPRC00,Fang07-iso-JPG}, which does not include the system evolution. In this article, the antisymmetric molecular dynamics (AMD) model is adopted to simulate the 140$A$ MeV $^{58, 64}$Ni +  $^9$Be reactions, and the simulated fragments will be analyzed using  the IBD method. The article is organized as follows. The IBD method and the AMD simulations are described briefly in Sec. \ref{model}. The IBD results for the simulated reactions are discussed in Sec. \ref{RAD} and a summary is presented in  Sec. \ref{summary}.

\section{methods description}
\label{model}

\subsection{IBD probe}
\label{ibdintro}
In canonical ensembles theory within the grand canonical limit, the cross section of a fragment is expressed as \cite{GrandCan,Tsang07BET}
\begin{equation}\label{yieldGC}
\sigma(I, A) = CA^{\tau}exp\{[-F(I, A) + \mu_{n}N + \mu_{p}Z]/T\},
\end{equation}
where $C$ is a constant; $I\equiv N - Z$ is the neutron-excess; $T$ is temperature; $\mu_n$ ($\mu_p$) is the chemical potential of neutrons (protons), which depends on the the density and temperature of the system; and $F(I, A)$ is the free energy of fragment, which also depends on the temperature.
The IYR differing by 2 units in $I$ is defined as \cite{Huang10},
\begin{equation}\label{ratiodef}
R(I + 2, I, A) = \sigma(I + 2, A)/\sigma(I, A).
\end{equation}
Considering two reactions where the measurement situations are the same (where the temperature of the reactions can be assumed as the same), the IYR difference between the reactions, i.e., IBD, is defined as \cite{IBD13PRC,IBD13JPG,IBD14Ca,IBDCa48EPJA},
\begin{eqnarray}\label{IBDIS}
\Delta\mu/T&=\mbox{ln}[R_{2}(I + 2, I, A)] - \mbox{ln}[R_{1}(I + 2, I, A)], \nonumber\\
&=(\Delta\mu_{n21} - \Delta\mu_{p21})/T, \hspace{2.8cm}\nonumber\\
&=[(\mu_{n2} - \mu_{n1}) - (\mu_{p2} - \mu_{p1})]/T, \hspace{1.5cm}
\end{eqnarray}
with the indices 1 and 2 denoting the reaction systems. $\Delta\mu/T$ is related to the density difference between neutrons and protons of the reaction systems \cite{IBD14Ca,IBDCa48EPJA}. Although the IBD probe is deduced from the canonical ensemble theory, the modified Fisher model can yield the same form of the IBD probe \cite{Huang10,ModelFisher3,MFM1}.

\subsection{AMD simulations}
\label{amdintro}
As one of the most sophisticated transport models, AMD describes the nuclear reaction at the microscopic level of interactions of individual nucleons \cite{AMD96,AMD99,AMD03,AMD04}. The extended version of AMD (AMD-V) introduces the wave-packet diffusion effect as a new quantum branching process and calculates the wave-packet diffusion effect with the Vlasov equation, which can predict the excitation energies of fragments better than other microscopic models \cite{AMD99}. For a complete description of the AMD model and the fragment analysis, the readers are referred to the more original references \cite{AMD96,AMD99,AMD03,AMD04,AMD04Wada,Huang10,Mocko08-AMD}. In this article, the AMD-V version is used to simulate the fragment produced in the 140$A$ MeV $^{58, 64}$Ni + $^9$Be reactions.  The standard Gogny (Gogny-g0) interaction \cite{Gogny-g0} is used to take into account all reaction processes. More than $10^{5}$ events for the 140$A$ MeV $^{58, 64}$Ni + $^9$Be reactions are treated using the AMD model. In the fragment analysis, a coalescence algorithm is adopted with a relative small coalescence radius $R_c =$ 2.5 fm in the phase space at $t =$ 500 fm/c. The primary fragments recognized in the phase space of the AMD simulation are allowed to decay by the sequential decay code GEMINI \cite{GEMINI}. To study the effect of central and peripheral collisions, two limitations on the impact parameters are adopted in the fragment analysis, i.e., $b1 =$ 0 -- 2  fm, and $b2$ = 0 -- 9 fm. In the previous work carried out by M. Mocko \textit{et al.}, the AMD simulations were set for an impact parameter range of 0-10 fm and up to the time of 150 fm/c by adopting the Gogny-AS interaction \cite{Mocko08-AMD}.

\section{results and discussion}
\label{RAD}

The isotopic distributions in the 140$A$ MeV $^{58, 64}$Ni + $^9$Be reactions calculated by the SAA, AMD, AMD + GEMINI and EPAX2/EPAX3 models have been compared in our recent work \cite{Ma15CPL,MaCW09PRC}. The AMD and AMD + GEMINI simulations can reproduce the cross sections for the symmetric fragments, but overestimate the cross sections of the neutron-rich fragments. In this work, we first compare the results of fragments with the same neutron-excess $I$.

\begin{figure}[htbp]
\includegraphics
[width=8.cm]{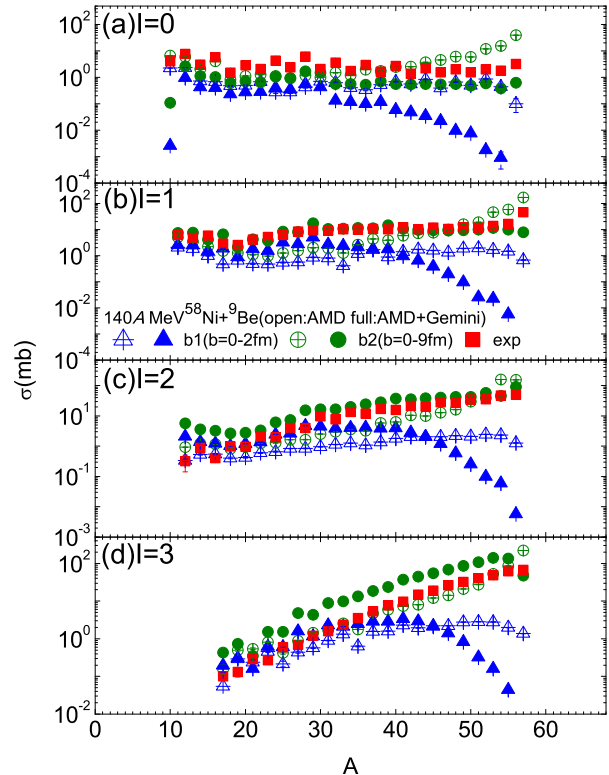}
\caption{\label{IBCSNi58} (Color online) The cross section distributions
of fragments with $I$ from 0 to 3, which are produced in the 140$A$ MeV $^{58}$Ni + $^9$Be reactions. The triangles
and circles denote the calculated cross sections with parameter ranges $b1$ ($b$ = 0 -- 2 fm)
and $b2$ ($b$ = 0 -- 9 fm), respectively. The open and full symbols denote the results for
the AMD and AMD + GEMINI calculations. The measured cross sections of fragments  \cite{Mocko06}
are plotted as squares.
}
\end{figure}

The distributions for fragments with $I$ from 0 to 3 in the 140$A$ MeV $^{58}$Ni + $^9$Be reaction are plotted in Fig. \ref{IBCSNi58}. The measured fragment distributions for the $I =$ 0 and 1 chains form plateaus. While for the $I > 1$ chains the measured fragment distribution increases with $A$. The calculated distributions for $I > 1$ fragments by AMD with the $b2$ limitations decrease with the increasing $A$ when $A < \sim 15$ and then form plateaus (or increase) with $A$. Compared to the measured fragments, the calculated fragments by AMD with $b1$ limitation show similar distributions. For fragments from $I = $ 0 to 3 chains, the calculated cross sections by AMD + GEMINI with the $b2$ limitation change from underestimating to overestimating the measured ones, with good reproduction of the measured cross sections for the $I = $ 1 chain. Besides, the distributions are similar for the small--$A$ fragments within the $b1$ and $b2$ limitations, indicating that the small--$A$ fragments are mainly produced in the central collisions. For fragments with larger $A$, the cross section for AMD + GEMINI with the $b2$ limitation decreases suddenly with the increasing $A$ because the hot fragments in AMD cannot survive and decay to smaller ones. The sudden change of the cross sections of large--$A$ fragments in the central collisions has also been observed in previous works, which is explained as one phenomena of skin effects. \cite{MaCW10PRC,MACW10JPG,MaCW13CPC,Ma13finite}.

\begin{figure}[htbp]
\includegraphics
[width=8.cm]{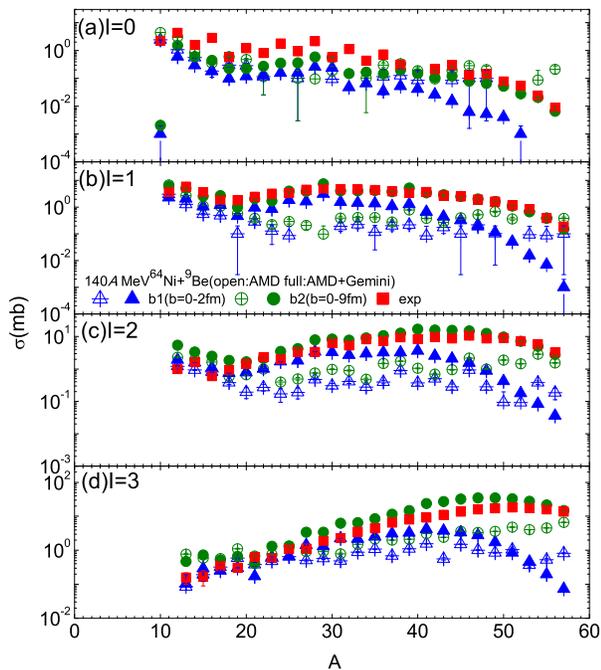}
\caption{\label{IBCSNi64} (Color online) The cross sectional distributions of fragments from $I =$ 0 to 3, which are produced in the 140$A$ MeV $^{64}$Ni + $^9$Be reactions. The triangles and circles denote the calculated cross sections with impact parameter ranges $b1$ ($b =$ 0 -- 2 fm) and $b2$ ($b =$ 0 -- 9 fm), respectively. The open and full symbols denote the results for the AMD and AMD + GEMINI calculations. The measured results by Mocko \textit{et al} \cite{Mocko06} are plotted as squares.
}
\end{figure}

The fragment distributions in the 140$A$ MeV $^{64}$Ni + $^9$Be reaction are plotted in Fig. \ref{IBCSNi64}. The plateau phenomena in the fragment  distributions in the $^{58}$Ni reaction is weakened in the $^{64}$Ni reaction. The results by AMD + GEMINI with the $b2$ limitation overestimate the measured one for the $I =$ 0 chain, but well reproduce the measured results for the $I =$ 1, 2, and 3 chains. The cross sections by AMD with the $b1$ and $b2$ limitations are similar for most of the fragments except those with $A$ close to the projectile nucleus, indicating that the cross sections for these fragments are slightly influenced by the impact parameters. The cross sections calculated by AMD + GEMINI with the $b1$ and $b2$ limitations are also similar when $A  < \sim 30$, which also indicates that they are less influenced by the impact parameters compared to the $A > \sim 30$ fragments. Besides, in both of the $^{58}$Ni and $^{64}$Ni reactions,  an obvious odd--even staggering phenomena is shown in the distribution for the $I =$ 0 chain.

\begin{figure}[htbp]
\includegraphics
[width=8.cm]{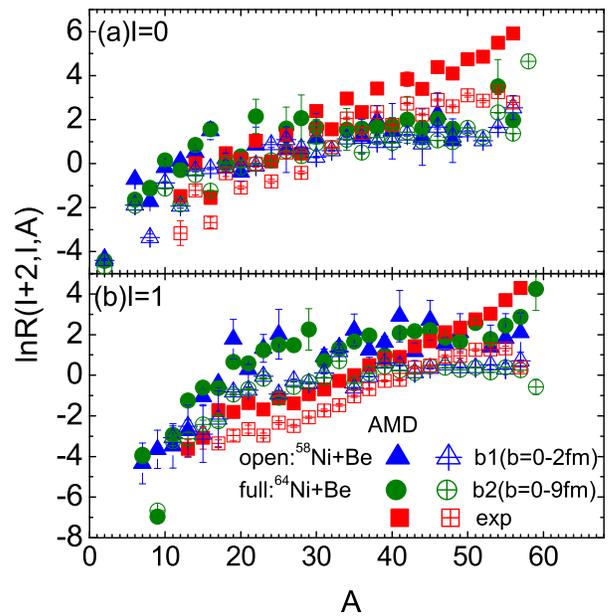}
\caption{\label{IYR-pre-I01} (Color online) The isobaric yield ratio for the $I =$ 0 and 1 chains in the 140$A$ MeV $^{58, 64}$Ni + $^9$Be reactions. The measured results are plotted as squares. The triangles and circles denote the calculated results by AMD with the impact parameter limitations $b1$ (0 -- 2 fm) and $b2$ (0 -- 9 fm), respectively. The open and full symbols denote the results for the $^{58}$Ni and $^{64}$Ni reactions, respectively.
}
\end{figure}

In Fig. \ref{IYR-pre-I01}, the  $I = $ 0 and 1 IYRs for the $^{58}$Ni and $^{64}$Ni reactions simulated by the AMD model are plotted.
The IYRs with the $b1$ and $b2$ limitations are very similar both for the $^{58}$Ni and $^{64}$Ni reactions in the AMD and AMD + GEMINI simulations, indicating that the IYRs for the $I = $0 and 1 chains are only slightly influenced by the impact parameters. The experimental IYRs increase with $A$, while in the AMD and AMD + GEMINI simulations, the IYR only increases with $A$ when $A < 30$ and it tends to be constant when $A > 30$. Given the obvious underestimation of the $I = $0 chain and the overestimation of the $I =$ 3 chain in the $^{58}$Ni reaction, the calculated IYRs by AMD do not agree with the measured ones very well. The same results are also shown in the AMD results for the $^{64}$Ni reaction. In the measured and calculated results with $b2$ limitation for the $^{58}$Ni reaction, the increasing IYRs are suppressed in the large--$A$ fragments. The obvious even-odd staggering is found in the experimental IYR distributions of $I =$ 0 chain for both the $^{58}$Ni and $^{64}$Ni reactions.

\begin{figure}[htbp]
\includegraphics
[width=8.cm]{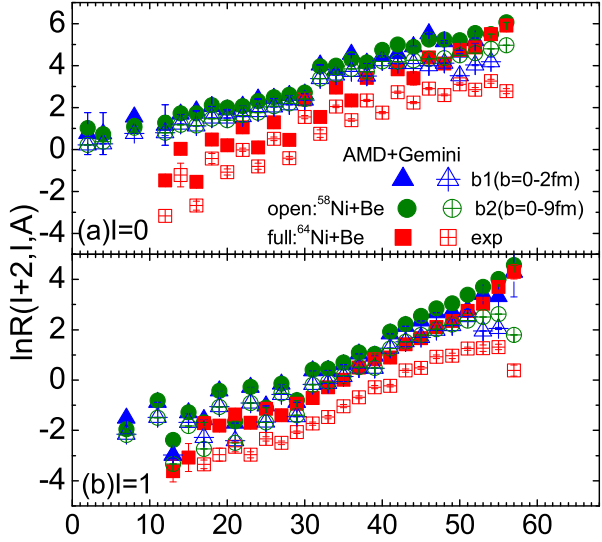}
\caption{\label{IYR-final-I01} (Color online) The isobaric yield ratio for the $I =$ 0 and 1 chains in the 140$A$ MeV
$^{58, 64}$Ni + $^9$Be reactions. The isobaric yield ratio for the measured results are plotted
as squares. The triangles and circles denote the AMD + GEMINI simulated results with impact
parameter ranges $b1$ (0 -- 2 fm) and $b2$ (0 -- 9 fm), respectively. The open and full symbols
denote the results for the $^{58}$Ni and $^{64}$Ni reactions, respectively.
}
\end{figure}

The IYRs for the $I = $ 0 and 1 chains calculated by AMD + GEMINI are plotted in Fig. \ref{IYR-final-I01}. For both the $^{58}$Ni and $^{64}$Ni reactions, after the decayed process through GEMINI, the IYRs with the $b1$ and $b2$ limitations overlap for most of the fragments, which means that after the decay, the IYRs for the $I = $0 and 1 chains are only slightly influenced by the impact parameters. The IYRs increase with $A$ of the fragments, which is different than those of the AMD results. The AMD + GEMINI calculated IYRs overestimate the measured results for both the $^{58}$Ni and $^{64}$Ni reactions, respectively. The suppression of IYRs for the large-$A$ fragments in the $^{58}$Ni reaction can be also found in the measured and calculated results with the $b2$ limitation. 

\begin{figure}[htbp]
\includegraphics
[width=8.cm]{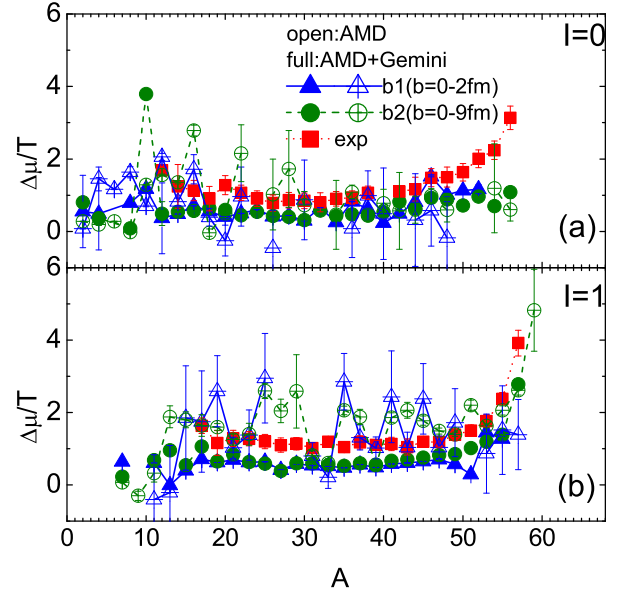}
\caption{\label{IBDcomp} (Color online) The isobaric yield ratio difference (IBD) for the
$I =$ 0 [in panel (a)] and $I =$ 1 [in panel (b)] chains between the 140$A$ MeV $^{58, 64}$Ni + $^9$Be
reactions. The IBDs for the measured results are plotted as squares. The triangles and circles
denote the calculated results with impact parameter ranges $b1$ (0 -- 2 fm) and $b2$ (0 -- 9 fm),
respectively. The open and full symbols denote the results for the AMD and AMD + GEMINI simulations, respectively.
}
\end{figure}

The IBD results between the 140$A$ MeV $^{58, 64}$Ni + $^{9}$Be reactions are plotted in Fig. \ref{IBDcomp}. When $A < 40$ and $A < 53$, the measured IYRs for the $I = $0 and 1 chains form quite good plateaus around $\Delta\mu/T \sim 1.0$, respectively.  In the AMD simulated results with the $b1$ and $b2$ limitations, the IBDs tend to be similar, but a slight difference is shown in the distribution (some of the IBD results are not shown due to the absence of cross sections in calculation). In the AMD results with the $b2$ limitation, for the $I = $ 0 chain, in trend the IBD is smaller than that with the $b1$ limitation. This phenomena is more clearly seen in the $I =$ 1 chain. In the AMD + GEMINI simulations with the $b1$ and $b2$ limitations, the IBDs for the $I = $ 0 and 1 chains almost overlap, indicating that for them the central collisions almost determine the IBD results. Thus the IBDs for the $I = $0 and 1 chains reflect the nuclear density in the central collisions. The measured IBD results for the $I =$ 0 and 1 chains are slightly underestimated by the AMD + GEMINI results, and overestimated by the AMD results. It is indicated that the AMD + GEMINI simulations can better reproduce the experimental IBD distributions. In addition, the calculated IBD results with the $b1$ and $b2$ limitations reveal that the increasing part of the IBD distribution is because of the different trends of IYRs for the peripheral reactions of $^{58}$Ni and $^{64}$Ni. An evident staggering is found in the IBDs for the AMD simulated reactions, but disappears in the IBDs for the measured and the AMD + GEMINI simulated reactions. From Fig. \ref{IYR-pre-I01}, it can be seen that the staggering in the IYRs for AMD simulated reactions for $^{58}$Ni and $^{64}$Ni occurs in a slightly different manners. In the measured IYRs and the IYRs for the AMD + GEMINI simulated reactions for the $^{58}$Ni and $^{64}$Ni reactions, the staggering occurs in a similar manner.

The result of $\Delta\mu/T$ has been related to the density difference between neutrons and protons for the reaction systems \cite{IBDCa48EPJA,NST2015IBD}, which is $\Delta\mu/T = \mbox{ln}(\rho_{n2}/\rho_{p2})-\mbox{ln}(\rho_{n1}/\rho_{p1})$. For the $^{58}$Ni reaction, if $\rho_{n1}/\rho_{p1} = $ 1 can be assumed \cite{IBD13PRC},  $\Delta\mu/T = \mbox{ln}\rho_{n2}-\mbox{ln}\rho_{p2} \equiv \Delta\mbox{ln}\rho_{np}$ for the asymmetric $^{64}$Ni system can be obtained. According to the IBD plateaus for the $I = $0 and 1 chains, in the $^{64}$Ni + $^9$Be reaction, $\Delta\mu/T \approx 1$ will result in $\Delta\mbox{ln}\rho_{np} = 1$ for the measured reaction, and $\Delta\mu/T \approx 0.5$ will result in $\Delta\mbox{ln}\rho_{np} = 0.65$ for the AMD + GEMINI simulated reaction. The $\Delta\mu/T$ for the $p$ + Kr and $p$ + Xe reactions have been estimated to be around 1.16 \cite{MFM1}. The IBD plateaus for the measured 140$A$ MeV $^{48}$Ca/$^{48}$Ca + $^{9}$Be reactions \cite{Mocko06} have been estimated to be around 1.85 $\pm$ 0.25 \cite{IBD13PRC,info2,NST2015IBD}. Assuming that for the $^{40}$Ca reaction $\rho_n/\rho_p = $1, for the  $^{48}$Ca reaction, one has $\Delta\mbox{ln}\rho_{np} = 1.85$. A relationship between $\Delta\mu/T$ and $\Delta\rho_{np}$ for the calcium reactions has been roughly shown \cite{IBDCa48EPJA}, in which $\Delta\mu/T = $ 1.85, 1, and 0.5 correspond to $\Delta\rho_{np}$ of 0.015, 0.009, and 0.004 fm$^{-3}$, respectively.

\section{summary}
\label{summary}
The AMD (+ GEMINI) models have been used to simulate the measured fragments in the 140$A$ MeV $^{58, 64}$Ni + $^{9}$Be reactions. The cross sections of fragments in the simulated reactions are analyzed by adopting two limitations on the impact parameters, i.e., $b1 = 0 - 2 $ fm and $b2 = 0 - 9$ fm, which reflect the central collisions and the whole reaction system. With a difference between the IBD results for the AMD and AMD + GEMINI simulations, they can reproduce the trend of the experimental IBDs for the $I = $ 0 and 1 chains. It is concluded that for the $I =$ 0 and 1 chains, the IBD plateaus are mainly determined by the central collisions, which reflects the density difference in the core of the reaction system. The increasing part of the IBD distribution is revealed to be the difference of neutron and proton densities between the peripheral reactions of the systems since the IYRs in the peripheral reactions of the symmetric system are suppressed. From the IBD results, it is concluded that the AMD + GEMINI simulation can better reproduce the measured ones. 

\begin{acknowledgments}
This work is supported by the Program for Science and Technology Innovation Talents in Universities of Henan Province (13HASTIT046), China. We are thankful for the helpful guidance from M.-R. Huang for the AMD simulation and the fragment analysis. The AMD simulation was performed on the High-Performance Computing Center at the College of Physics and Electrical Engineering, HNU.
\end{acknowledgments}

\end{document}